\documentclass[12pt]{article}

\usepackage{bbm}
\usepackage{epsfig}
\usepackage{array}
\usepackage{float}
\usepackage{dsfont}
\usepackage{amstext}
\usepackage{rotating}
\usepackage{a4}
\usepackage{a4wide}
\usepackage{cite}
\usepackage{graphicx}
\usepackage{amsmath} 

\newcommand{\diag}{\mathrm{diag}}
\newcommand{\matrixx}[1]{\begin{pmatrix} #1 \end{pmatrix}}

\def\be{\begin{equation}}
\def\ee{\end{equation}}
\def\gs{\mathrel{
   \rlap{\raise 0.511ex \hbox{$>$}}{\lower 0.511ex \hbox{$\sim$}}}}
\def\ls{\mathrel{
   \rlap{\raise 0.511ex \hbox{$<$}}{\lower 0.511ex \hbox{$\sim$}}}}

\newcommand{\ba}{\begin{array}{c}}
\newcommand{\baz}{\begin{array}{cc}}
\newcommand{\barrr}{\begin{array}{rrr}}
\newcommand{\bad}{\begin{array}{ccc}}
\newcommand{\bav}{\begin{array}{cccc}}
\newcommand{\baf}{\begin{array}{ccccc}}
\newcommand{\bea}{\begin{equation} \begin{array}{c}}
\newcommand{\eea}{ \end{array} \end{equation}}
\newcommand{\ea}{\end{array}}
\newcommand{\D}{\displaystyle}
\newcommand{\dms}{\mbox{$\Delta m^2_{\odot}$}}
\newcommand{\dma}{\mbox{$\Delta m^2_{\rm A}$}}

\newcommand{\gsim}{\raise0.3ex\hbox{$\;>$\kern-0.75em\raise-1.1ex\hbox{
   $\sim\;$}}} 
\newcommand{\lsim}{\raise0.3ex\hbox{$\;<$\kern-0.75em\raise-1.1ex\hbox{
   $\sim\;$}}}

\begin{document}

\title{\Large \bf
Relating large $U_{e3}$ to the ratio of neutrino mass-squared
differences }

\author{
Werner Rodejohann$^a$\thanks{email: 
\tt werner.rodejohann@mpi-hd.mpg.de}~,~~
Morimitsu Tanimoto$^b$\thanks{email: 
\tt tanimoto@muse.sc.niigata-u.ac.jp}~,~~
Atsushi Watanabe$^{a}$\thanks{email: 
\tt atsushi.watanabe@mpi-hd.mpg.de}
\\\\
{\normalsize \it $^a$Max--Planck--Institut f\"ur Kernphysik,}\\
{\normalsize \it  Postfach 103980, D--69029 Heidelberg, Germany}\\ \\ 
{\normalsize \it $^b$Department of Physics, Niigata University,}\\
{\normalsize \it Niigata 950--2181, Japan} }
\date{}
\maketitle
\thispagestyle{empty}
\vspace{0.8cm}
\begin{abstract}
\noindent  
The non-zero and sizable value of $U_{e3}$ puts pressure on 
flavor symmetry models which predict an initially vanishing value. 
Hence, the tradition of relating fermion mixing matrix elements with 
fermion mass ratios might need to be resurrected. 
We note that the recently observed non-vanishing value of $U_{e3}$ can
be related numerically to the ratio of solar and atmospheric
mass-squared differences. The most straightforward realization of
this can be achieved with a combination of texture zeros and a vanishing neutrino
mass. We analyze the implications of some of these possibilities and
construct explicit flavor symmetry models that predict these
features. 

\end{abstract}

\newpage
\section{\label{sec:intro}Introduction}
Neutrino physics has entered again an exciting period. The upper limit
on the last unknown lepton mixing angle, $\theta_{13}$, was almost
unchanged since the Chooz bound was released in 1999
\cite{hep-ex/9907037}. After the first weak hints towards a non-zero
value of this important parameter appeared, see the early analysis in
\cite{arXiv:0806.2649}, more and more evidence supporting $\theta_{13}
\neq 0$ was accumulated, as demonstrated in
Refs.~\cite{arXiv:1001.4524,arXiv:1103.0734,arXiv:1106.6028,arXiv:1108.1376,arXiv:1111.3330}.
The case of vanishing $\theta_{13}$ was (almost) closed during the
last year by results from the T2K \cite{t2k}, MINOS
\cite{minos} and Double Chooz \cite{dc} experiments, and 
combined analyses are showing evidence for $|U_{e3}| = \sin
\theta_{13}$ exceeding the $3\sigma$ level. For instance, 
Ref.~\cite{arXiv:1111.3330} finds at the $1.96\sigma$ level that 
\be \label{eq:ue3_range}
|U_{e3}|_{\rm nor} = 0.144_{-0.068}^{+0.061} ~\mbox{ and }~
|U_{e3}|_{\rm inv} = 0.149_{-0.067}^{+0.062} \, , 
\ee
for the normal and inverted ordering, respectively. An analysis of 
T2K, MINOS and Double Chooz data gave (for the normal ordering) the $3\sigma$ range
\cite{dc} 
\be \label{eq:thomas}
|U_{e3}| = 0.146_{-0.119}^{+0.084} \, . 
\ee
While being very probably non-zero, $\theta_{13}$ remains of course the smallest
lepton mixing angle. Usually, lepton mixing is described
mainly by tri-bimaximal mixing, or other mixing schemes with
$U_{e3}=0$. The motivation here is that the smallness of $U_{e3}$ is 
attributed to the presence of a flavor symmetry which predicts it to be
zero. In such models, the masses (eigenvalues of mass matrices) are
independent of mixing angles (eigenvectors of mass matrices). 
See Refs.~\cite{rev1,rev2} for recent reviews on flavor symmetry
models. While corrections leading to sizable values of $U_{e3}$ are 
possible in flavor symmetry models, and are in fact analyzed
frequently, usually all mixing angles receive corrections of
the same order. While $\theta_{12}$ lies, according to observations,
very close to its tri-bimaximal value $\sin^2 \theta_{12} = \frac 13$,
and $\theta_{23}$ is very well compatible with maximal mixing, the
sizable value of $|U_{e3}| \simeq 0.14 $ implies a particular
perturbation structure, which seems somewhat tuned or put in by hand.

\begin{figure}[t]
	\begin{center}
\scalebox{1.3}{
		\includegraphics[width=0.45\textwidth]{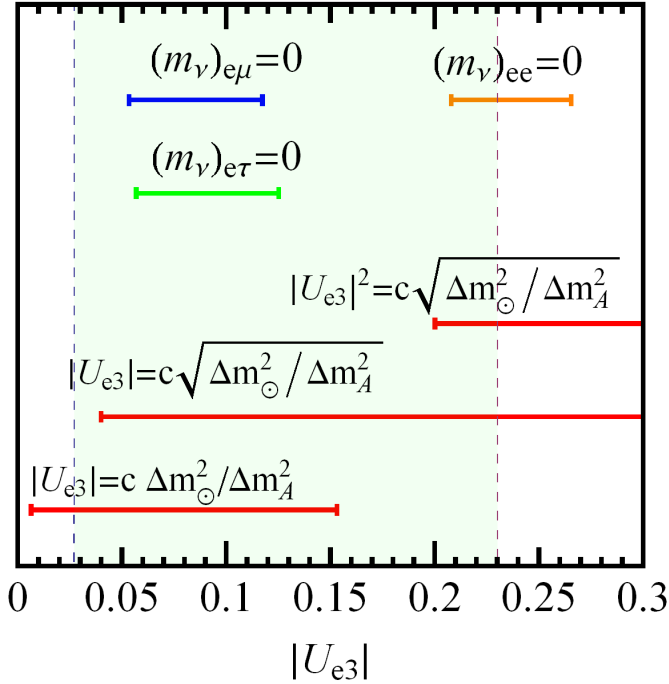}

}
	\end{center}
		\caption{Allowed ranges of $|U_{e3}|$ for the three 
relations reproduced here in explicit models. The shaded area is the
current $3\sigma$ range of $|U_{e3}|$. 
Also shown are the currently allowed $3\sigma$ ranges corresponding to $|U_{e3}| = c \,
\sqrt{\dms/\dma}$, $|U_{e3}|^2 = c \, \sqrt{\dms/\dma}$ and $|U_{e3}| = c \,
\dms/\dma$, where $c$ is varied between 0.25 and 4. Note that the
relation $|U_{e3}| = c \,\dms/\dma$ is not considered in this paper,
but given here for completeness as it also reproduces the data very well.}
	\label{fig:1}
\end{figure}

At this point, it is worth to recall the Gatto-Sartori-Tonin
relation $\sin \theta_C \simeq \sqrt{m_d/m_s}$ \cite{53130}, which
links the Cabibbo angle to a quark mass ratio. Such intriguing
relations between fermion mass ratios and mixing matrix elements were 
in the past driving forces for approaches to study the flavor
problem. 
Motivated by the sizable value of $|U_{e3}|$ and the moderate 
neutrino mass hierarchy as implied by the comparably large ratio of 
the neutrino mass-squared differences, we attempt in this note to connect
those two quantities. As a byproduct, the two small quantities in
neutrino physics, $|U_{e3}|$ and the ratio of mass-squared
differences, are linked. Indeed, the $3\sigma$ range of 
$|U_{e3}|^2 \simeq 0.001 - 0.053$ lies order-of-magnitude-wise close to the 
ratio of the solar ($\dms$) and atmospheric ($\dma$) mass-squared
differences\footnote{We will apply here and in what follows the
$3\sigma$ ranges of $|U_{e3}|$ from Eq.~(\ref{eq:thomas}) and of the
analysis from Ref.~\cite{arXiv:1108.1376} for the remaining
oscillation parameters.}, 
$\dms/\dma \simeq  0.026- 0.038$, or $\sqrt{\dms/\dma} \simeq
0.160- 0.196$. 
In a three-flavor framework, which is necessary to
consider when $U_{e3}$ is involved, one can expect that $|U_{e3}|^2 =
c \, \sqrt{\dms/\dma}$, $|U_{e3}| = c \, \sqrt{\dms/\dma}$ 
or $|U_{e3}| = c \, \dms/\dma$, where 
$c$ is an order one number and function of the other
mixing angles. This can easily lead to agreement even with the central
value of $|U_{e3}|$, as illustrated in Fig.~\ref{fig:1}. Note 
that the relative uncertainty on $|U_{e3}|$ is currently around 100 \%. 
 
We propose here very simple and straightforward realizations of the
observations made above, namely 
\be \label{eq:main1}
|U_{e3}|^2 = \sqrt{\frac{\dms}{\dma}} \, \sin^2 \theta_{12} = 
0.054_{-0.011}^{+0.016}\, , 
\ee
and 
\bea \label{eq:main2} \D 
|U_{e3}| \simeq  \frac 12 \, 
\sqrt{\frac{\dms}{\dma}} \, \sin 2 \theta_{12} \cot \theta_{23} = 
0.078_{-0.025}^{+0.040}\, , \\ \D 
 |U_{e3}| \simeq  \frac 12 \, 
\sqrt{\frac{\dms}{\dma}} \, \sin 2 \theta_{12} \tan \theta_{23}= 
0.084_{-0.027}^{+0.041} \, .
\eea
These relations are obtained by setting in the normal mass ordering
the smallest neutrino mass to zero and by asking the $ee$, $e\mu$ or $e\tau$
element of the neutrino mass matrix in the flavor basis to vanish. We
stress that other possibilities for similar relations surely exist,
but here we focus on these very simple ones. Our main motivation is to
note the potential link of ratios of fermion masses and $U_{e3}$ as an
alternative approach in model building.

In what follows we will analyze the predictions of these relations 
and present simple flavor symmetry models, based on the discrete
groups $S_3$ and
$D_4$, respectively, which reproduce them. As mentioned above, we will obtain our relations by
combining the single texture zero approach \cite{arXiv:hep-ph/0603111}
in the flavor basis with the case of a vanishing neutrino mass \cite{hep-ph/0212341}. The 
relations we will obtain are simple, and have of course been present in the literature before,
see for instance Ref.~\cite{arXiv:1108.4010}. However, as far as we
know they were neither presented with
the motivation that we outlined above, nor with any underlying flavor
symmetry model input. \\

The outline of the paper is as follows: in Section \ref{sec:general}
we will analyze the relations (\ref{eq:main1}) and (\ref{eq:main2}) in
a model-independent way.  Simple models predicting them are discussed
in Section \ref{sec:models}, before we conclude in Section
\ref{sec:concl}. Some necessary flavor group details are delegated to
the Appendix.

\section{\label{sec:general}General Analysis}

The lepton mixing matrix can be parameterized as 
\begin{align}\nonumber
\begin{split}
	U &=  
\matrixx{c_{12} c_{13} & s_{12} c_{13} & s_{13} e^{-i\delta}\\
	-c_{23} s_{12}- s_{23} s_{13} c_{12} e^{i\delta} & c_{23} c_{12}- s_{23} s_{13} s_{12} e^{i\delta} & s_{23} c_{13}\\
	s_{23}s_{12}- c_{23} s_{13} c_{12} e^{i\delta} & -s_{23} c_{12}- c_{23} s_{13} s_{12} e^{i\delta} & c_{23} c_{13}}  
\diag (1,  e^{i\alpha_1}, e^{i\alpha_2})\,. 
\end{split}
\label{eq:U}
\end{align}
Note that we will work in this section in the flavor basis, i.e.~the
charged lepton mass matrix is diagonal. The neutrino mass matrix is given by 
\be \label{eq:mnu}
m_\nu = U \, \diag (m_1, m_2, m_3) \, U^T \, . 
\ee
In the normal hierarchy case, setting the smallest neutrino mass $m_1$
to zero and asking the $\alpha \beta$ entry of the mass matrix to vanish, corresponds to the relation
\be
U_{\alpha 2} \, U_{\beta 2} \, m_2 + U_{\alpha 3} \, U_{\beta 3} \,
m_3 = 0 \, . 
\ee
This in turn gives two relations which describe the phenomenological
results of the scenario, one for the absolute value 
\be \label{eq:abs}
\frac{\left|U_{\alpha 3} \, U_{\beta 3} \right| }{\left|U_{\alpha 2}
\, U_{\beta 2} \right|} = \frac{m_2}{m_3} = \sqrt{\frac{\dms}{\dma}}
\, , 
\ee
and one for the phases
\be \label{eq:phase}
{\rm arg}(U_{\alpha 3} \,  U_{\beta 3} \,  U_{\alpha 2}^\ast \, U_{\beta
2}^\ast) = \pi  \, . 
\ee
The first relation (\ref{eq:abs}) will give a constraint on the value
of $|U_{e3}|$, the second relation (\ref{eq:phase}) can relate the two
physical CP phases with each other. Note that with $m_1 = 0$ only one  
Majorana phase is present. Simple
modifications of the above relations can be made in case the inverted
hierarchy is considered, but for the sake of brevity we will not give
the relevant expressions here. 

Consider now the case of setting the $ee$ element of the neutrino mass
matrix to zero. With $\alpha = \beta = e$, the result from
Eq.~(\ref{eq:abs}) is 
\be 
\tan^2 \theta_{13} = \sqrt{\frac{\dms}{\dma}} \, \sin^2 \theta_{12} \,, 
\ee
which with $\tan^2 \theta_{13}\simeq |U_{e3}|^2 \, (1 + |U_{e3}|^2 )$
corresponds to our Eq.~(\ref{eq:main1}). The relevant exact expression
for the phases from Eq.~(\ref{eq:phase}) is 
\be
\alpha_1 - \alpha_2 = \frac 12 \pi - \delta \, . 
\ee
The allowed range of $|U_{e3}|$ with this scenario is given in
Fig.~\ref{fig:1}. Note that with the $ee$ element of the mass matrix being
zero, there will be no contribution to neutrino-less double beta decay
from light neutrinos \cite{Rodejohann:2011mu}. 

\begin{figure}[t]
	\begin{center}
\scalebox{1.5}{
		\includegraphics[width=0.45\textwidth]{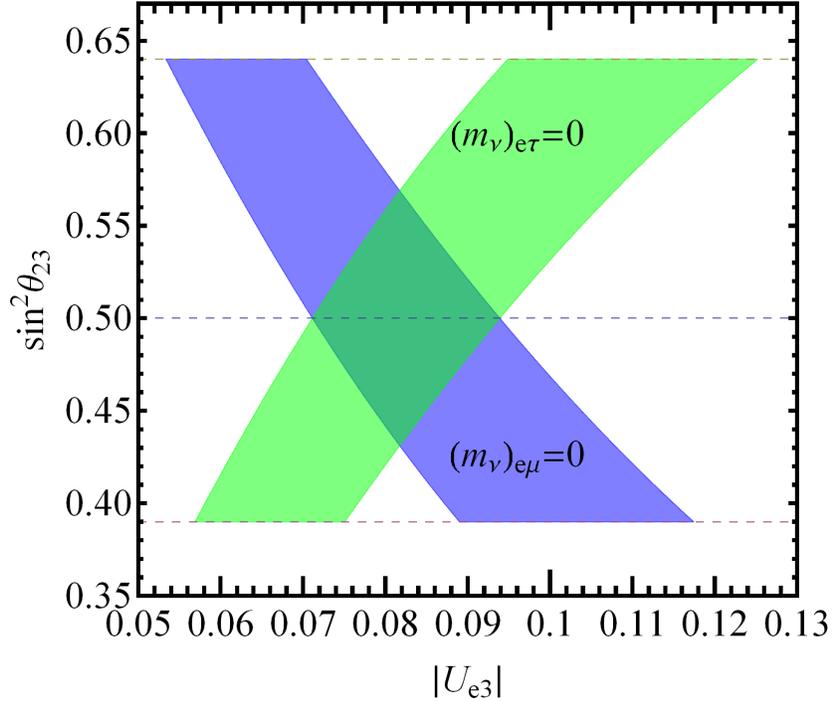}
}
	\end{center}
		\caption{The dependence of $|U_{e3}|$ on $\sin^2
\theta_{23}$ for the cases of vanishing $m_1$ and a zero $e\mu$
($e\tau$) element of the mass matrix.}
	\label{fig:2}
\end{figure}

The next case is when we set the $e\mu$ element of the neutrino mass
matrix to zero. This gives at leading order in $|U_{e3}|$ the already
quoted result from Eq.~(\ref{eq:main2}): 
\be
|U_{e3}| \simeq \frac 12 \, \sqrt{\frac{\dms}{\dma}} \, \sin 2
\theta_{12} \cot \theta_{23} \, , 
\ee
and furthermore, again at leading order,  
\be
\alpha_1 - \alpha_2 \simeq \frac 12 (\pi - \delta) \, . 
\ee
The third case occurs for a vanishing $e\tau$ element of $m_\nu$, for
which we get the same result as for a vanishing $e\mu$ element, with
the replacement $\cot \theta_{23} \to \tan \theta_{23}$ and 
\be
\alpha_1 - \alpha_2 \simeq -\frac 12 \delta \, . 
\ee
The predictions for $|U_{e3}|$ are shown in Fig.~\ref{fig:1}. The
dependence on the atmospheric neutrino parameter $\sin^2 \theta_{23}$
is displayed in Fig.~\ref{fig:2}. Note that there is no dependence on
this parameter when the $ee$ element is zero.

Let us remark that the predictions are very stable under corrections
of renormalization group running.\\

In principle our analysis could be extended to cases for which the
remaining entries of the mass matrix are zero. It is easy to see that
if the smallest mass $m_1$ in the normal hierarchy is zero, the
$\mu\mu$, $\mu\tau$ and $\tau\tau$ elements cannot vanish. In the
inverted hierarchy case with $m_3 = 0$, the $ee$ element cannot be
zero. The remaining possibilities in the inverted hierarchy suffer
from little predictivity and do not link the ratio of mass-squared
differences and $U_{e3}$ in a straightforward manner.  
For instance, if we would set in the
inverted hierarchy case with $m_3=0$ the $e\mu$ element of the mass
matrix to zero, we would get 
\[ 
|U_{e3}| \, \cos \delta \simeq \frac 14 \frac{\dms}{\dma} \frac{\sin 2
\theta_{12}}{\tan \theta_{23}} \, . 
\]
A similar relation holds for the $\mu\tau$ block of $m_\nu$, setting
for instance the $\mu\mu$ entry to zero yields 
\[ 
|U_{e3}| \, \cos \delta \simeq \frac 14 \frac{\dms}{\dma} \frac{\cos
2\theta_{12} \, \tan \theta_{12}}{\cos^2 \theta_{12} \, \tan \theta_{23} } \, . 
\]
This can be traced back to the fact that when $m_3 = 0$ the only
remaining masses are $m_2 \simeq m_1$, i.e.~they do not possess a
hierarchy and do not allow to make a straightforward and direct
relation between $|U_{e3}|$ and a small ratio, simply because there is
no small ratio of masses.

\section{\label{sec:models}Simple Model Realizations}
In this section we present several examples of flavor models which produce the
desired features of the relations in Eqs.~(\ref{eq:main1}, \ref{eq:main2}). 
The key ingredients to get these relations are the vanishing elements in the 
Majorana mass matrix $m_\nu$ and one massless state in the active
neutrino spectrum. The latter property is naturally explained by a seesaw mechanism
with two right-handed neutrinos \cite{FGY}, see \cite{Guo:2006qa} for
a review. We attribute the former property to texture zeros in the Dirac and
the Majorana mass matrices in the original Lagrangian at some high-energy scale.
The texture zeros are realized along the line of~\cite{HY,S3tex}, where the discrete
flavor symmetries and their breakdown by new scalar fields play the central role.

\subsection{A Model for $(m_\nu)_{ee} =0$}
Let us first discuss a model which produces Eq.~(\ref{eq:main1}).
We introduce $S_3 \times Z_3$ as a discrete flavor symmetry and
require the presence of scalar fields which carry nontrivial charges
of the flavor symmetry. 
The particle content relevant for lepton masses and mixings are
summarized in the following table: 
\begin{center}
\renewcommand{\arraystretch}{1.25}
\begin{tabular}{ccccccc|cc}\hline\hline
      &$(\overline{L_1},\overline{L_2})$&$\overline{L_3}$&$(\nu_{R_1},\nu_{R_2})$&$e_R$ 
& $\mu_R$ & $\tau_R$ & $(\phi_1,\phi_2)$
& $\chi$\\\hline
$S_3$ & $2^*$& $1_{\rm S}$ &$2$ & $1_{\rm S}$ & $1_{\rm S}$ & 
$1_{\rm S}$ & $2$ & $1_{\rm S}$\\
$Z_3$ & $\omega$ & $\omega$ & $\omega$ & $\omega$ & $1$ & $\omega^2$ & 
$\omega$ & $\omega$ \\\hline
\end{tabular}
\end{center}

\noindent Here $\omega = e^{2\pi i/3}$, $L_i$ are the left-handed lepton doublets, $\nu_{R_i}$ are the right-handed 
neutrinos, $e_R,\mu_R,\tau_R$ are the right-handed charged leptons. 
The scalar fields $\phi_{1,2}$ and $\chi$ are so-called flavons, which are singlet
under the Standard Model gauge group and whose vacuum expectation value (VEV)
break the flavor symmetry.
Details for the tensor products of $S_3$  
are presented in Appendix~\ref{appendixA}. 

At the energy scales where the flavor symmetries are unbroken,
the neutrino Yukawa interactions and the Majorana masses 
for $\nu_{R_i}$ are written in terms of higher-dimensional operators which involve
the Standard Model fields and the flavons.
At the leading order of the inverse power of the cutoff scale,
they are given by
\begin{eqnarray} 
-\mathcal{L}_\nu &\,=\,&\nonumber
y_1 \,(\overline{L_1}\nu_{R_1} + \overline{L_2}\nu_{R_2} )\chi H^*\,+\,
y_2 \,(\overline{L_1}\nu_{R_2} \phi_2 + \overline{L_2}\nu_{R_1}\phi_1 )H^*\\
&&\,+\, y_3 \,\overline{L_3}(\nu_{R_1} \phi_2 + \nu_{R_2}\phi_1 )H^*
 \,+\, \frac{1}{2}g_1(\overline{\nu_{R_1}^c}\nu_{R_1}\phi_1 +  \overline{\nu_{R_2}^c}\nu_{R_2}\phi_2) \\
&&\,+\, \frac{1}{2}g_2(\overline{\nu_{R_1}^c}\nu_{R_2} +  \overline{\nu_{R_2}^c}\nu_{R_1})\chi \,+\, {\rm h.c.},\nonumber
\end{eqnarray}
where $H$ is the standard model Higgs field, $y_i$ are coupling constants 
which carry inverse mass dimension, while $g_i$ are dimensionless constants. 
After $\phi_i$ and $\chi$ obtain vacuum expectation values according
to the alignment 
\begin{eqnarray}
\phi_i \to 
\begin{pmatrix}
\langle \phi_1 \rangle\\
0\\
\end{pmatrix},
\quad
\chi \to \langle \chi \rangle,
\label{vac1}
\end{eqnarray}
and electroweak symmetry breaking $H \to \langle H \rangle = (v,0)^{\rm T}$, 
the mass matrices at low energy take the forms
\begin{eqnarray}
m_D \simeq \begin{pmatrix}
a & 0 \\
b & a \\
0 & c\\
\end{pmatrix},\quad
M_R \simeq \begin{pmatrix}
M_A & M_B \\
M_B & 0 \\
\end{pmatrix},
\label{mDMR1}
\end{eqnarray}
where $a = y_1 \langle \chi \rangle v$, 
$b = y_2 \langle \phi_1 \rangle v$, $c = y_3 \langle \phi_1 \rangle v$,
$M_A = g_1 \langle \phi_1 \rangle $ and $M_B = g_2 \langle \chi \rangle$.
By assuming $m_D \ll M_R$ and performing the seesaw diagonalization, one
finds the effective Majorana mass matrix 
\begin{eqnarray}
m_\nu \,\simeq\,
\begin{pmatrix}
0 & a^2 & ac \\
a^2& 2ab & bc \\
ac & bc & 0 \\
\end{pmatrix}
\frac{1}{M_B}
\,- \,
\begin{pmatrix}
0 & 0 &0 \\
0& a^2 & ac \\
0 & ac & c^2 \\
\end{pmatrix}
\frac{M_A}{M_B^2} \, .
\label{mnu1}
\end{eqnarray}
The charged leptons are diagonal in this model (see below). 
The vanishing of $(m_\nu)_{ee}$ is thus realized due to the texture zeros in
the Dirac and the Majorana mass matrices. 
Notice that the texture~(\ref{mDMR1}) is equivalent to one of the patterns
discussed in~\cite{texture} and its physical implications are already studied.
Central values of the mass-squared differences and the solar and
atmospheric mixing angles are obtained for instance by setting 
$a/\sqrt{M_B} \simeq 0.079\,{\rm eV^{1/2}}$, $b/\sqrt{M_B} \simeq 0.013\,{\rm eV^{1/2}}$
with the conditions $c \simeq b$, $M_A \simeq -M_B$. 
Since our purpose here is to present an example for a model leading to
$(m_\nu)_{ee} = 0$, we do not discuss (\ref{mnu1}) further.\\

The charged lepton sector is written by a combination of
higher-dimensional operators and a 
renormalizable operator. At leading order of the inverse of the
cutoff, it is given by
\begin{eqnarray} \nonumber
-\mathcal{L}_l &\,=\,& 
y_e \,(\overline{L_1}\phi_1 + \overline{L_2}\phi_2 )e_R H,+\,
y_\mu \,(\overline{L_1}\phi_2^* + \overline{L_2}\phi_1^* )\mu_R H
\,+\, y_\tau \,\overline{L_3}\,\tau_R H \,+\,{\rm h.c.},
\end{eqnarray}
where $y_e$ and $y_\mu$ are parameters of inverse mass dimension while
$y_\tau$ is dimensionless. 
In the vacuum Eq.~(\ref{vac1}), the charged lepton mass matrix is diagonal 
\begin{eqnarray}
M_l \simeq \begin{pmatrix}
y_e \langle \phi_1 \rangle & 0 & 0 \\
0 & y_\mu \langle \phi_1 \rangle^* & 0 \\ 
0 & 0 & y_\tau 
\end{pmatrix}v.
\label{ml1}
\end{eqnarray}
Due to the effective nature of the muon mass term, the hierarchy
between $m_\mu$ and $m_\tau$ is naturally explained. The smaller electron mass
can easily be explained by adding an additional $U(1)$ symmetry under
which the right-handed electron field is charged.\\

For the realization of desired flavor structure, the ``asymmetric''
VEV configuration $\phi_i \to (\langle \phi_1 \rangle, 0 )^{\rm T}$  
in Eq.~(\ref{vac1}) 
is playing a vital role. Such an alignment has been shown to be easily
possible in generic potentials in $S_3$ theories in
Ref.~\cite{HY}. Furthermore, in Refs.~\cite{twist} it was shown to
achieve the required alignment via boundary conditions of scalar 
fields in extra-dimensional space, which force either of the two
components to have a zero mode.

\subsection{Models for $(m_\nu)_{e\mu} =0$ and $(m_\nu)_{e\tau} =0$}
Next we present a model for Eq.~(\ref{eq:main2}).
We assume $S_3\times Z_4$ flavor symmetry and introduce gauge singlet
scalars $\xi_1, \xi_2$ and $\eta$. The charge assignment is as
follows: 
\begin{center}
\renewcommand{\arraystretch}{1.25}
\begin{tabular}{cccccccc|cc}\hline\hline
      &$(\overline{L_1},\overline{L_2})$&$\overline{L_3}$&$\nu_{R_1}$&$\nu_{R_2}$ & $e_R$ 
& $\mu_R$ & $\tau_R$ & $(\xi_1,\xi_2)$
& $\eta$\\\hline
$S_3$ & $2$& $1_{\rm S}$ &$1_{\rm S}$ & $1_{\rm S}$ & $1_{\rm A}$ & $1_{\rm A}$ & 
$1_{\rm S}$ & $2$ & $1_{\rm S}$\\
$Z_4$ & $i$ & $i$ & $1$ & $-1$ & $1$ & $-1$ & $-i$ & 
$i$ & $i$ \\\hline
\end{tabular}
\end{center}

\noindent 
At leading order of the inverse of the cutoff scale,
the flavor-symmetric neutrino Yukawa Lagrangian is written as
\begin{eqnarray}
-\mathcal{L}_\nu &\,=\,&
z_1 \,(\overline{L_1}\xi_1^* + \overline{L_2}\xi_2^* )\nu_{R_1}H^*\,+\,
z_2 \,(\overline{L_1}\xi_2 + \overline{L_2}\xi_1 )\nu_{R_2}H^* \nonumber\\
&&\,+\, z_3 \,\overline{L_3}\,\nu_{R_1}\eta^* H^*\,+\,
z_4 \,\overline{L_3}\,\nu_{R_2}\eta H^* \\
&& \,+\, \frac{1}{2}(M_1 \overline{\nu_{R_1}^c}\nu_{R_1} + M_2 \overline{\nu_{R_2}^c}\nu_{R_2}) \,+\, {\rm h.c.}, \nonumber
\end{eqnarray}
where $z_i$ are coupling constants 
which carry inverse mass dimension and $M_{1,2}$ are the Majorana masses for
the right-handed neutrinos.
After the doublet $(\xi_1,\xi_2)$ and the singlet $\eta$ develop the vacuum expectation values
\begin{eqnarray}
\xi_i \to 
\begin{pmatrix}
\langle \xi_1 \rangle\\
0\\
\end{pmatrix},
\quad
\eta \to \langle \eta \rangle,
\label{vac2}
\end{eqnarray}
and the electroweak symmetry breaks down, 
the neutrino mass matrices are given by
\begin{eqnarray}
m_D \simeq \begin{pmatrix}
p & 0 \\
0 & q \\
r & s\\
\end{pmatrix},\quad
M_R \simeq \begin{pmatrix}
M_1 & 0 \\
0 & M_2 \\
\end{pmatrix},
\label{mDMR2}
\end{eqnarray}
where $p = z_1 \langle \xi_1 \rangle^* v$, 
$q = z_2 \langle \xi_1 \rangle v$, $r = z_3 \langle \eta \rangle^* v$,
and $s = z_4 \langle \eta \rangle v$.
If $p,q,r,s \ll M_{1,2}$ so that the seesaw mechanism works,
the mass matrix for the left-handed neutrinos reads
\begin{eqnarray}
m_\nu \,\simeq\,
\begin{pmatrix}
p^2 & 0 & pr \\
0& 0 & 0 \\
pr & 0 & r^2 \\
\end{pmatrix}
\frac{1}{M_1}
\,+\,
\begin{pmatrix}
0 & 0 &0 \\
0& q^2 & qs \\
0 & qs & s^2 \\
\end{pmatrix}
\frac{1}{M_2} \, .
\end{eqnarray}
The charged lepton mass matrix is again diagonal, and is just as in
the previous subsection given by a combination of renormalizable terms
for the tau lepton mass and effective terms for the electron and muon
mass. The vanishing of $(m_\nu)_{e\mu}$ is achieved by the texture zeros in the Dirac
and the Majorana mass matrices~(\ref{mDMR2}). We note that such
textures have been discussed in \cite{Goswami:2009bd}.

With the simple replacement $L_2 \leftrightarrow L_3$ one can modify the model to
generate $(m_\nu)_{e\tau} = 0$. \\

\newpage

Another model to obtain a vanishing $(m_\nu)_{e\mu}$ uses the flavor group
$D_4 \times Z_2$: 
\begin{center}
\renewcommand{\arraystretch}{1.25}
\begin{tabular}{cccccc|ccccc}\hline\hline
      &$(\overline{L_1},\overline{L_2})$&$\overline{L_3}$&$(\nu_{R_1},\nu_{R_2})$ & $(e_R,\mu_R)$ 
&  $\tau_R$ & $\eta_1^-$ & $\eta_2^+$  & $\eta_3^+$  & $\eta_4^-$ &
$(\xi_1^+,\xi_2^+)$  \\\hline 
$D_4$ & $2$ & $1_1$ & $2$ & $2$ & $1_1$ & $1_1$ & $1_2$ & $1_3$ &
$1_4$ & $2$ \\ 
$ Z_2$ & $+$ & $+$ & $+$ & $-$ & $+$ & $-$ & $+$ & $+$ & $-$ & $+$ \\ \hline
\end{tabular}
\end{center}

\noindent 
With this identification and the multiplication rules in the
convention of Ref.~\cite{Hagedorn:2005kz} (see Appendix \ref{appendixB}), it is straightforward to
see that the charged lepton and the right-handed neutrino mass
matrices are diagonal, while the Dirac mass matrix has a texture as in
(\ref{mDMR2}). In contrast to the $S_3$ model, there is no VEV
alignment necessary, at the price however of introducing 6 weak
doublets $\eta_1^-$, $\eta_2^+$, $\eta_3^+$, $\eta_4^-$ and
$(\xi_1^+,\xi_2^+)$.


\section{\label{sec:concl}Summary and Conclusions}

We stressed here that the recently emerging non-zero $\theta_{13}$
puts some pressure on models with an initially vanishing value, and 
that it may be natural that the two small but non-zero quantities in
neutrino oscillations, $U_{e3}$ and the ratio of mass-squared
differences, are linked. Indeed, the ratio of mass-squared 
differences is numerically close to the value of $U_{e3}$. 
The most straightforward application of this idea leads to 
the relations $|U_{e3}|^2 =  \sqrt{\dms/\dma} \, \sin^2
\theta_{12} \simeq 0.05$ and  
$|U_{e3}| \simeq  \frac 12 \, 
\sqrt{\dms/\dma} \, \sin 2 \theta_{12}  \simeq 0.08$, 
 in good agreement with data. There may be other realizations
of this and similar observations, and interesting model building
opportunities, somewhat alternative to the usual considerations, may arise.

\vspace{0.3cm}
\begin{center}
{\bf Acknowledgments}
\end{center}
WR is supported by the ERC under the Starting Grant 
MANITOP and by the DFG in the Transregio 27, MT is supported 
by the Grand-in-Aid for Scientific Research No.~21340055, and AW 
by the Young Researcher Overseas Visits Program for Vitalizing Brain
Circulation Japanese in JSPS No.~R2209.

\newpage
\appendix
\section{Group Details}
For the sake of completeness, we give here the necessary details to
reproduce the models in Section \ref{sec:models}. The complete group
structure can be found in Ref.~\cite{HY} for $S_3$ and
\cite{Hagedorn:2005kz} for $D_4$. 

\subsection{The group $S_3$}
\label{appendixA}
We use here the complex representation of $S_3$, see for instance 
Ref.~\cite{HY}. With $S_3$ doublets $\psi = \begin{pmatrix}\psi_1 \\ \psi_2 \\ \end{pmatrix}$ and 
$\phi = \begin{pmatrix}\phi_1 \\ \phi_2 \\ \end{pmatrix}$, $2\times 2$
is decomposed as 
\begin{eqnarray} \nonumber 
\psi^* \times \phi = 
\underbrace{\begin{pmatrix} \psi_1^* \phi_2 \\ \psi_2^* \phi_1\\ \end{pmatrix}}_{2}
+ \underbrace{(\psi_1^* \phi_1 -\psi_2^* \phi_2) }_{1_{\rm A}}
+ \underbrace{(\psi_1^* \phi_1 +\psi_2^* \phi_2) }_{1_{\rm S}} \, ,
\end{eqnarray}
and
\begin{eqnarray}\nonumber 
\psi \times \phi = 
\underbrace{\begin{pmatrix} \psi_2 \phi_2 \\ \psi_1 \phi_1\\ \end{pmatrix}}_{2}
+ \underbrace{(\psi_1 \phi_2 -\psi_2 \phi_1) }_{1_{\rm A}}
+ \underbrace{(\psi_1 \phi_2 +\psi_2 \phi_1) }_{1_{\rm S}} \, .
\end{eqnarray}

\subsection{The group $D_4$}
\label{appendixB}
In the convention used here, the four singlets of $D_4$ multiply as
follows: 
\begin{center}
\begin{tabular}{c|cccc}
$\times$&$1_1$&$1_2$&$1_3$&$1_4$\\
\hline
$1_1$        &$1_1$       &$1_2$       &$1_3$       &$1_4$\\
$1_2$        &$1_2$       &$1_1$       &$1_4$       &$1_3$\\
$1_3$        &$1_3$       &$1_4$       &$1_1$       &$1_2$\\
$1_4$        &$1_4$       &$1_3$       &$1_2$       &$1_1$\\
\end{tabular}
\end{center}
Two doublets $(\psi_1, \psi_2)^{\rm T}$ and $(\phi_1, \phi_2)^{\rm T}$
multiply as $2 \times 2 = 1_1 + 1_2 + 1_3 + 1_4$, where 
 \[
(\psi_1 \phi_1 + \psi_2 \phi_2)/\sqrt{2} \sim 1_1 \; , 
 \;\;\;  (\psi_1 \phi_2 + \psi_2 \phi_1)/\sqrt{2} \sim 1_2 \, ,
\]
\[
(\psi_1 \phi_2 -\psi_2 \phi_1)/\sqrt{2} \sim 1_3 \; , 
 \;\;\;  (\psi_1 \phi_1 -\psi_2 \phi_2)/\sqrt{2} \sim 1_4 \,.
\]


\end{document}